\documentclass[%
 reprint,
 amsmath,amssymb,
 aps,
]{revtex4-1}

\usepackage{graphicx}
\usepackage{dcolumn}
\usepackage{bm}
\usepackage{amsfonts}
\usepackage{amssymb}
\usepackage[latin1]{inputenc}
\usepackage{textcomp}
\usepackage{epstopdf}

\begin{document}



\title{Bloch-like super-oscillations and unidirectional motion of phase driven quantum walkers}

\author{A. R. C. Buarque, M. L. Lyra and W. S. Dias}
\address{Instituto de F\' isica, Universidade Federal de Alagoas, 57072-900 Macei\' o, Alagoas, Brazil.}

\begin{abstract}
We study the dynamics of a quantum walker simultaneously subjected to time-independent and -dependent phases. Such dynamics emulates a charged quantum particle in a lattice subjected to a superposition of static and harmonic electric fields. With proper settings, we investigate the possibility to induce Bloch-like super-oscillations, resulting from a close tuning of the frequency of the harmonic phase $\omega$ and that associated with the regular Bloch-like oscillations $\omega_B$. By exploring the frequency spectra of the wavepacket centroid, we are able to distinguish  the regimes on which regular and super-Bloch oscillations are predominant. Furthermore, we show that under exact resonant conditions $\omega=\omega_B$ unidirectional motion is established with the wavepacket average velocity being a function of the quantum walk coin operator parameter, the relative strengths of the static and harmonic terms, as well as the own phase of the harmonic phase. We show that the average drift velocity can be well described within a continuous-time analogous model.  
\end{abstract}

\maketitle

\section{introduction}
\label{introduction}

The classical random walk and its stochastic movement of particles form a well-known concept in physics, whose application extends to a wide variety of systems such as behavioral macroeconomics~\cite{NELSON1982139}, image segmentation~\cite{1704833}, animal dynamics~\cite{Viswanathan1996,Codling2008}, computer science~\cite{10.1007/11569596_31}, evolutionary ecology~\cite{Dieckmann1996,METZ1992198} and thermal conductivity of nanofluids~\cite{KEBLINSKI2002855,doi:10.1063/1.1756684}. Within the quantum context, discrete-time quantum walks (DTQWs) have received an outstanding attention in recent years. In general lines, its dynamics concerns a quantum walker whose movement through the lattice is closely related to its internal state, which changes step by step~\cite{aharonov1993quantum}. Coherent superposition and quantum interference make DTQWs usually faster than their classic counterpart, and therefore an interesting and versatile tool for the realization of quantum algorithms and quantum simulations~\cite{aharonov1993quantum,PhysRevLett.102.180501,PhysRevA.67.052307}. 

Due to its promising character, DTQWs have been explored on different environments, such as linear optics~\cite{su2019experimental,wang2018dynamic,PhysRevLett.106.180403}, electrodynamics cavities~\cite{sanders2003quantum}, Bose-Einstein condensates~\cite{PhysRevLett.104.100503,PhysRevLett.103.090504} and integrated photonic waveguides~\cite{grafe2020integrated,Crespi2013}. Furthermore, studies in DTQWs have examined different ingredients such as disorder~\cite{PhysRevLett.106.180403,Crespi2013,PhysRevA.89.042307,PhysRevE.100.032106}, nonlinearity~\cite{SR_nlqw,PhysRevA.101.023802}, particle-particle interaction~\cite{PhysRevA.83.042317,Carson2015} and non-Hermiticity~\cite{PhysRevA.93.062116,Longhi:19}. Here, we will focus on quantum walks whose addition of specific phases emulates the action of electric fields~\cite{PhysRevLett.111.160601,genske2013electric,bru2016electric,PhysRevLett.118.130501}. Considering a phase which depends linearly on position, Cedzich {\it et al.} described the long-time propagation properties of quantum walker as very sensitive to the emulated electric field. Ballistic and localized quantum walks have been reported by employing rational and irrational phases, respectively~\cite{PhysRevLett.111.160601}. An experimental realization of discrete time quantum walks which simulate the effect of an electric field on a charged particle has been reported in Ref.~\cite{genske2013electric}. Using single Cs atoms in spin-dependent optical lattices and measures by fluorescence imaging, it has been shown that a quantum particle can exhibit features closely related to Bloch oscillations. By studying the phenomenology for a 2D system, Bru {\it et al.} reported that the particle dynamics is easily affected by orientation of the field~\cite{bru2016electric}. The analysis of conical intersections in the dispersion relations suggests suitable directions of the field for a perfect 2D trapping to occur. 

A description analogous to Bloch oscillations has also been described by considering time-dependent phases on both single- and split-step DTQWs protocols~\cite{PhysRevLett.118.130501}. The refocusing behavior has been described as resulting from the interplay between dynamical and geometric phases, as well as non-adiabatic transitions. Other studies exploring DTQWs with time-dependent phases have described dynamical localization and quasi-periodic dynamics~\cite{PhysRevA.73.062304,PhysRevA.93.032329,PhysRevLett.114.140502}.

Previous results showed how time-independent or -dependent phases have been employed on DTQWs to obtain dynamical localization and/or Bloch oscillations.  Here, our goal is to address to the following question: How DTQWs behave with the simultaneous presence of both ingredients? 
Such system corresponds to a quantum walker subject to the concomitant action of two artificial electric fields. The constant field is emulated by a phase with a linear dependence on position, while the time-dependent field comes from a phase which has a harmonic modulation. 
We show different quantum walk dynamics owing to the tuning of the electric fields characteristics. We observe the development of Bloch-like super-oscillations. Such oscillatory dynamics displays characteristic frequencies,  which can be adjusted by controlling the artificial electric fields. The crossover from regular to super-Bloch oscillations is established. Furthermore, we demonstrate that, under resonant conditions, the walker can develop a unidirectional drift to a preferential side of the lattice. Both drift velocity and direction of movement are explicitly shown to depend on the electric fields settings and the specific coin operator. Its overall behavior is shown to be well described by an analogous continuous time approach.

\section{model}
\label{model}

The problem consist of a quantum walker moving in a one-dimensional discrete chain with $N$ sites under the influence of a superposed artificial DC and AC electric fields. Such walker consists of a qubit (two-state quantum system) with the internal degree of freedom (e.g., spin~\cite{aharonov1993quantum} or polarization~\cite{PhysRevLett.106.180403}). Its full Hilbert space is composed of two subspaces $H=H_{P}\otimes H_{C}$: $H_{P}$ denotes the position-space spanned by a set orthonormal vectors $\{|n\rangle_{P}$: $n \in \mathbb{N}^{*}\}$. $H_{C}$ is defined by two-dimensional coin-space spanned by orthogonal vectors {$|\uparrow\rangle_{C}=[1,0]^{T}$, $|\downarrow\rangle_{C}=[0,1]^{T}$},  associated with the internal degree of freedom of the particle which determines the direction of the walk motion.

In general, the evolution step of a discrete-time quantum walk consists in two operations: We start by applying the unitary operator quantum gate (well-known as quantum coin) $\hat{C}$ in the initial state of particle, followed by the conditional shift operator $\hat{S}$. Thereby, the state $|\Psi_{0}\rangle$ of the particle after $t$ steps can be described by $|\psi_{t}\rangle=\hat{W}_{0}...\hat{W}_{0}|\Psi_{0}\rangle=\prod\hat{W}_{0}|\Psi_{0}\rangle$. Here, $\hat{W}_{0}:=\hat{S}\cdot(\hat{C}\otimes\mathbb{I}_{P})$ describes the evolution operator of the walk, with $\mathbb{I}_{P}$ representing the identity matrix over the subspace of the positions. The conditional shift operator $\hat{S}$ has the form 
\begin{eqnarray}\label{shift_operator}
\hat{S}=|\uparrow\rangle\langle \uparrow|\otimes \hat{S}_{+}+ |\downarrow\rangle\langle \downarrow|\otimes \hat{S}_{-},
\end{eqnarray}
with the operators $\hat{S}_{\pm}=\sum_{x=1}^{N}|n\pm 1\rangle\langle n|$. The quantum gate $\hat{C}$ is an arbitrary $SU(2)$ unitary operator given by
\begin{eqnarray}\label{quantum_coin}
\hat{C}=\left(\begin{array}{cc}
\cos\theta & \sin\theta\\
\sin\theta & -\cos\theta
\end{array} \right),
\end{eqnarray}
where $\theta$ is an adjustable parameter which controls the variance of the probability distribution.

In order to simulate the electric field, we define an extra unitary operator\cite{genske2013electric}
\begin{eqnarray}\label{field_operator}
\hat{F}_{E}=\sum_{n}\exp(iG\hat{n})|n\rangle\langle n|\otimes\mathbb{I}_{C}.
\end{eqnarray}
$G$ is a function that represents the phase imprinted by the effective electric field that can be expressed in fractions of $2\pi/m$, with $m\in\mathbb{R}$. This operator only acts in the subspace of the positions $H_{P}$. 
By exploring the Eq.\eqref{field_operator}, we observe quantum-walks under the presence of electric fields with $m={1,2}$ showing a similar behavior to those without an electric field.

\begin{figure}
    \centering
    \includegraphics[width=7.5cm]{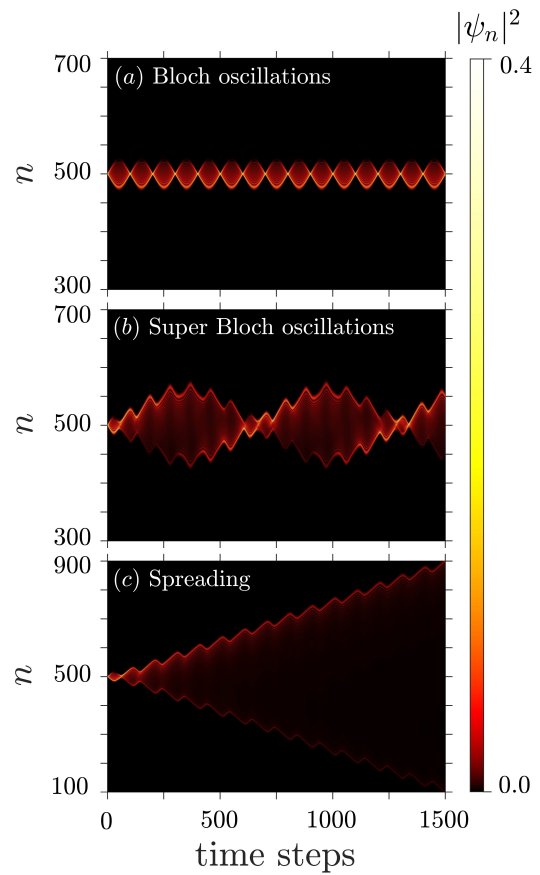}
    \caption{Probability density profile in the plane of positions $n$ vs time steps. (a) Under the presence of a uniform electric field $G$ with $m=100$ ($\Delta \Phi=0$), the quantum walker shows Bloch-like oscillations. (b) When subjected of both constant and harmonic fields, with $\omega$ very close to the Bloch frequency ($\omega_B$), the quantum walker performs oscillations with large amplitudes. (c) When we tune the frequency $\omega=\omega B$ the quantum walker spreads developing a preferential walk to one side of the chain.
}
    \label{fig1}
\end{figure}
In this work, we express $G$ so that it simulates the superposition between two electric fields: a uniform (DC) and a harmonic component (AC). Thus, we consider
\begin{eqnarray}\label{superposed_fields}
G(t)=\Phi_{0} + \Delta\Phi\sin(\omega t+\phi).
\end{eqnarray}
The first term represents the DC artificial field with magnitude $\Phi_{0}=2\pi/m$, while the second represents the AC artificial field with modulation $\Delta\Phi$, frequence $\omega$ and phase $\phi$. The time evolution of the system is now governed by the ``electrified'' operator, given by
$\hat{W}_{el}:=\hat{F}_{E}\cdot\hat{W}_{0}$. Thus, by using the time-evolution protocol $|\psi_{t}\rangle=\hat{W}_{el}|\psi_{t-1}\rangle$, we can derive the recursive evolution equations for the probability amplitudes
\begin{eqnarray}\label{recursive}
& &\psi_{t+1,n}^{\uparrow} = e^{iG_{t}(n-N/2)}
(\cos\theta \psi_{t,n-1}^{\uparrow} + \sin\theta \psi_{t,n-1}^{\downarrow}), \nonumber \\
& &\psi_{t+1,n}^{\downarrow} = e^{iG_{t}(n-N/2)}
(\sin\theta \psi_{t,n+1}^{\uparrow} - \cos\theta \psi_{t,n+1}^{\downarrow}),
\end{eqnarray}
with $\psi_{t,n}^{\uparrow}$ and $\psi_{t,n}^{\downarrow}$ representing the probability amplitudes of obtaining the states $|\uparrow\rangle$ and $|\downarrow\rangle$ at position $n$ and time-step $t$.

Although quantum walks have been experimentally realized in diverse physical systems, we consider the experimental implementation feasible by using optical feedback loop~\cite{PhysRevLett.106.180403} or integrated waveguide circuits~\cite{Crespi2013}, since such systems have been used to enforce time-dependent and time-dependent phases on the subspace of position of quantum walker. The first has an advantage of the demand for resources that remains constant as the number of steps in the quantum walk increases.

Our results were obtained following equations \eqref{recursive} for a quantum particle initially localized at lattice center $n_{0} = N/2$. We consider throughout the analysis open chains as the boundary condition, whose sizes $N$ are large enough that the wave function does not reach its edges during the evaluated time.

\section{Results and discussion}
\label{Results_and_discussion}  
 \begin{figure}
    \centering
    \includegraphics[width=\linewidth]{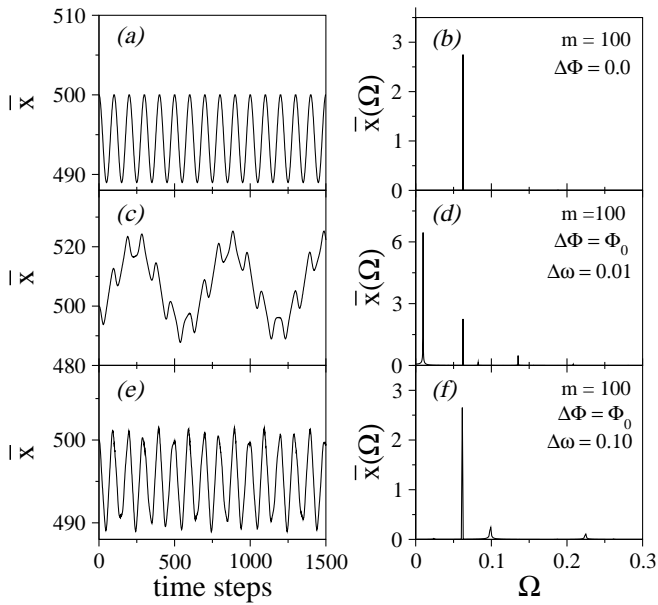}
    \caption{Left panels show the centroid of a quantum walker initially localized at lattice center $n_0=500$, with distinct configurations of $G(t)$. Right panels show their respective Fourier transform. (a) in absence of harmonic modulation $(\Delta\Phi=0)$, the centroid exhibits Bloch-like oscillations, whose frequency corresponds to magnitude of emulated field $\Phi_0=2\pi/m$. (b) By connecting the harmonic component and tuning its frequency to a value very close to the Bloch frequency ($\Delta\omega=\omega-\omega_B=0.01$), we observe large oscillations described as a superposition of two main
frequencies. (c) Oscillations with similar pattern of Bloch-like oscillations are recovered as we increase the detuning $\Delta\omega$.
    }
    \label{fig2}
\end{figure}
We start our discussion illustrating the time-evolution of the particle's quantum wavepacket under the effect of different configurations of artificial external fields. In Fig. \ref{fig1}, we show the profile of the probability density $|\psi_{n}|^{2}$ as a density plot in the plane of the positions $(n)$ vs time-steps for three representative cases of artificial fields. We consider the initial state of the quantum particle is a symmetric one of the form 
\begin{eqnarray}\label{initial_state}
|\Psi_{0}\rangle=1/\sqrt{2}(|\uparrow\rangle + i|\downarrow\rangle\otimes|n_{0}\rangle,
\end{eqnarray}
centered on $n_{0}=N/2$ with $N=10^{3}$ sites. We assume a weak artificial electric field with $m=100$ so that the uniform phase $G$ has increments of $\Phi_{0}=2\pi/100$. In addition, we use Hadamard gate ($\theta=\pi/4$) for which such initial wavepacket would evolve in time developing symmetric wavefronts. Revisiting the case of a quantum walker under the action of a constant phase having a linear dependence on the position,  the quantum walker performs oscillations around the initial position with a well-defined period and frequency, as  shown in Fig. \ref{fig1}(a). 
Thanks the presence $\hat{F}_{E}$ operator, such behavior is usually associated with Bloch Oscillations (BOs)~\cite{genske2013electric,PhysRevLett.118.130501,PhysRevA.101.062324}, an emerging phenomenon of solid state physics in which an electron is loaded in a periodic potential subjected to a constant electric field. This driving, which we treat as Bloch-like oscillations, is a quasi-stationary dynamics for a quantum walker that persists over a long-time whenever the phase increment $\Phi_{0}\ll 1$.
\begin{figure}
    \centering
    \includegraphics[width=6cm]{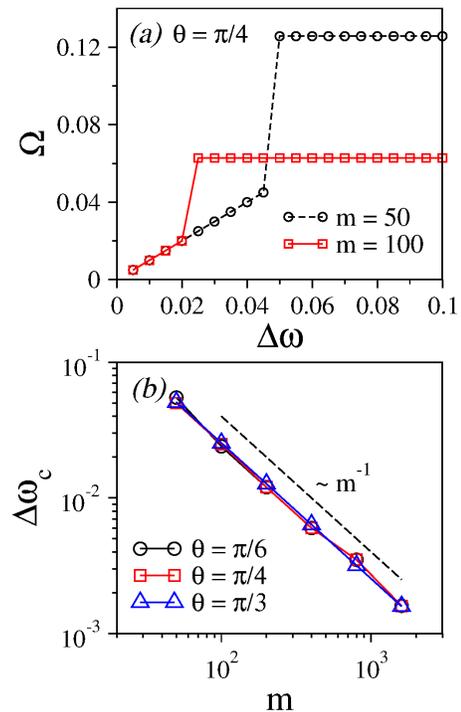}
    \caption{(a) Quantitative analysis of the predominant spectral frequency ($\Omega$) as a function of the detuning factor  ($\Delta\omega$) for distinct strengths of the emulated static field ($\Phi_{0}=2\pi/m$). We consider the quantum gate phase $\theta=\pi/4$. For small detuning, the predominant frequency is associated with Bloch-like super-oscillations. Above a characteristic $m$-dependent detuning, regular Bloch-like oscillations predominate. In (b) we show that the characteristic crossover frequency $(\Delta\omega_{c})$ is inversely proportional to $m$ and independent from applied quantum gate.
    }
    \label{fig3}
\end{figure}
In the very long-time regime, the oscillation become unstable~\cite{PhysRevLett.111.160601}. Now, we add a harmonic component ($\Delta\Phi=\Phi_0$) to the $G(t)$ phase that plays a role similar to an external time-dependent field, see Eq. (\ref{superposed_fields}). As shown in Fig. \ref{fig1}(b), a new dynamics emerges: the wave-function oscillates with large amplitudes in the position space when the frequency of the harmonic field component ($\omega$) is close to that of the Bloch-like oscillations ($\omega_{B}$). Analogous to solid-state super-Bloch oscillations, such phenomenon  arises only when the harmonic field frequency ($\omega$) is slightly detuned ($\omega-\omega_B=\Delta\omega\ll \Phi_{0}$) from the oscillation frequency $(\omega_{B})$. This scenario that we label as Bloch-like super-oscillations changes when we adjust the frequencies for the case of exact resonance. In Fig. \ref{fig1}(c) we assume $\Delta\omega\rightarrow 0$ so that, $\omega=\omega_{B}$. The wave-packet of the quantum walker develops a preferential walk direction, in close analogy to the dynamics of single electron wavepackets obeying a continuous-time Schr\"odinger equation under similar field conditions. 

For a more detailed description of the interplay between regular and super large amplitude Bloch oscillations, we compute the time evolution of the centroid of the particle's quantum wavepacket in order to understand the effects caused by the detuning $\Delta\omega$ on its dynamics. The centroid is defined as
\begin{eqnarray}\label{centroid}
\overline{\textrm{x}}(t)=\sum_{n} n|\psi_{n}(t)|^{2}.
\end{eqnarray}
In the left column of Fig. \ref{fig2}, we show the evolution of the average position of the particle $\overline{\textrm{x}}$ for three field settings $G(t)$. For all cases, the centroid exhibits an oscillatory pattern. However, changing the value of $\Delta\omega$ the dynamics visits two regimes. 
Fig. \ref{fig2}(a) displays the case where the quantum-particle is only under the action of a uniform artificial electric field, $\Delta\Phi=0.0$ and $\Phi_{0}=2\pi/100$. The centroid shows coherent oscillations with Bloch period $T_{B}=m$. The oscillation frequency is proportional to the acquired uniform phase, $\Phi_{0}$. The Fourier transform of the centroid over several oscillations $\overline{\textrm{x}}(\Omega)$, reported in Fig. \ref{fig2}(b), clearly shows that the centroid displays an oscillatory pattern with predominant frequency $\omega_{B}=\Phi_{0}$. The detailed dynamics depends closely on whether $\Phi_{0}/2\pi$ is an irrational or a rational number, as discussed in \cite{PhysRevLett.111.160601,genske2013electric,bru2016electric}.
\begin{figure}
    \centering
    \includegraphics[width=7.cm]{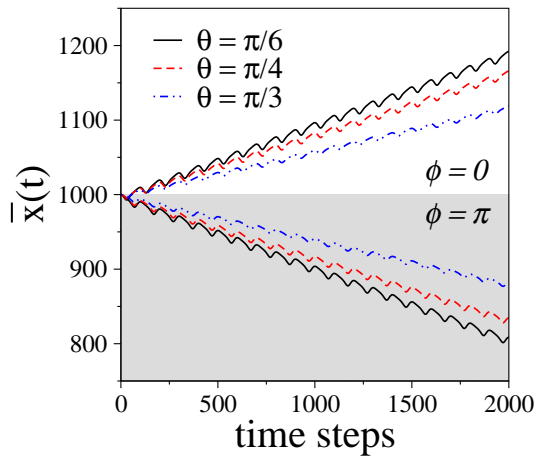}
    \caption{Time evolution of the quantum walker centroid for different quantum gates $\theta=\pi/6, \pi/4$ and $\pi/3$. Numerical experiments were performed using $m=100$, $\Delta\Phi=\Phi_{0}$ and $\omega=\omega_{B}$. We demonstrate the possibility of driving the quantum walker along the lattice (to the left or to the right) by controlling the parameter $\phi$.}
    \label{fig4}
\end{figure}
When we turn on the harmonic component of the artificial electric field with modulation amplitude $\Delta\Phi=\Phi_{0}$ and with a frequency very close to $\omega_{B}$, i.e,  $\omega=\Delta\omega+\omega_{B}$, the quantum particle develops a large amplitude oscillation  [see Fig. \ref{fig2}(c)],  with a period $T_{SBO}\approx 2\pi/\Delta\omega\gg T_{B}$. The spectral decomposition of the centroid trace shows that its dynamics has two main components [see In Fig. \ref{fig2}(d)].  The largest component  represents  the frequency of SBOs $(\Omega_{SBO})=\Delta\omega = 2\pi/m - \omega=0.01$. The second component is the usual Bloch frequency, which depends only on the uniform DC component $\Phi_{0}=2\pi/100$. It is also possible to identify a small third frequency component which is the sum of the frequencies $\omega_{B} + \omega + \Delta\omega$. 
It is interesting to mention that the dynamics of SBOs are independent of the phase $\phi$  of the artificial AC field. In Fig. \ref{fig2}(e), we consider a larger detuning by increasing the difference between the frequencies $\Delta\omega=0.1$. In this case, the influence of the harmonic component of the artificial electric field give rises just to small amplitude slow oscillations of the centroid. The predominant component of its spectral decomposition remains the standard Bloch frequency, as shown in Fig. \ref{fig2}(f).

The previous analysis allows us to identify two distinct dynamical regimes according to the detuning between the frequency of the harmonic artificial field and the frequency of the standard Bloch oscillations. These two regimes are characterized by the predominant spectral frequency $\Omega$ of the centroid trace. In Fig. \ref{fig3}(a), we report $\Omega$ as a function of $\Delta\omega$ for the Hadamard gate and two distinct phase increments $\Phi_{0}=2\pi/m$. For  low detuning values $(\Delta\omega\leq \Delta\omega_c)$, a linear dependence of the predominant frequency on $\Delta\omega$ signals that it is associated with the SBO. On the other hand, when $(\Delta\omega>\Delta\omega_c)$ the standard $m$-dependent Bloch frequency ($\Omega=2\pi/m$) predominates over the dynamics of the wavepacket. By considering a chain with $N=2500$, $t_{max}=40N$ time steps and three distinct quantum gates ($\theta=\pi/6, \pi/4, \pi/3$), we report in Fig. \ref{fig3}(b) the characteristic detuning frequency separating two dynamical regimes as a function of $m$. We show the critical detuning dependent with $1/m$, but independent from the quantum gate. Thus, we unveil the crossover frequency being proportional to the standard Bloch-like oscillations. Our data support $\Delta\omega_c/\omega_B \simeq 0.3979$.


\begin{figure}
    \centering
    \includegraphics[width=6.75cm]{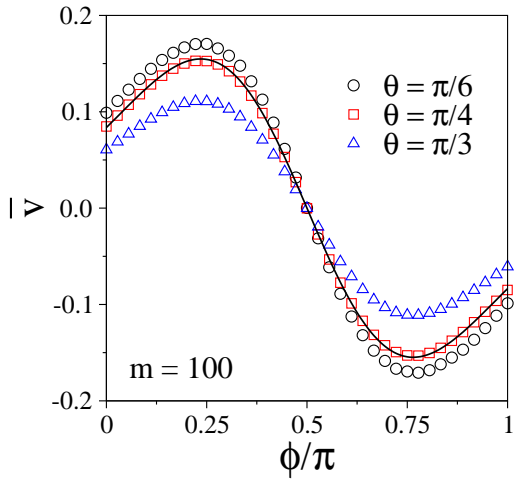}
    \caption{Phase dependence of the quantum walker centroid velocity for three distinct quantum gates: $\theta=\pi/6, \pi/4$ and $\pi/3$. Solid line corresponds to the continuous-time theoretical result given by equation \ref{eq:velocity}. }
    \label{fig5}
\end{figure}
In the absence of frequency detuning $\Delta\omega=0$, i.e., the frequency of the sinusoidal phase coinciding with the Bloch frequency $(\omega=\omega_{B})$, the quantum particle develops a preferential walk. For a more quantitative description, we compute numerically the centroid of the quantum walker wavepacket under resonant field conditions.
In Fig. \ref{fig4}, we plot the time-evolution of the average position of the quantum walker for three different quantum gate configurations, $\theta=\pi/6$ (continuous-line), $\pi/4$ (dashed) and $\pi/3$ (dashed-dotted). We used $m=100$. 
A unidirectional modulated walk of the wave-packet centroid is obtained whose average velocity depends on the relative phase of the AC field and of quantum gate parameter $\theta$. For $\phi=0$ the quantum particle performs a unidirectional walk to the left of the initial position. The opposite occurs when $\phi=\pi$.  This effect can be used to manipulate the dynamics of the quantum particle by tuning the phase of the artificial harmonic field.

In order to explore the unidirectional walk induced by the resonance condition, we determined the phase dependence of the centroid's velocity. Fig. \ref{fig5}
displays the average velocity $(\overline{\mathrm{v}})$ as a function of artificial harmonic field phase $\phi$. Here, we also consider the field amplitudes $\Phi_0=\Delta\Phi$. We unveil that under resonance condition associated to the two AC/DC field components, the quantum particle dynamics, initially in the quantum state Eq. \eqref{initial_state}, undergoes a unidirectional motion with a sinusoidal-like dependence of the average velocity on the phase of the AC component for all quantum gates. 
In the regime of small field increments $\Phi_0$, leading to Bloch oscillations with period $T_B\gg 1$, one expects a close analogy between the quantum walk dynamics and that of a single particle wavepacket governed by a continuous-time Schr\"odinger equation within a tight-binding approximation. It has been demonstrated that  the wave-packet
net velocity of a non-interacting particle driven by an AC field
in resonance with the BO is given by\cite{PhysRevA.83.053627,CAETANO20112770,doi:10.1002/pssb.201600805}
\begin{equation}
v\propto v_0\cos\left(\delta\Phi\cos(\phi)-\phi\right),
\label{eq:velocity}
\end{equation}
where $\delta\Phi=(\Delta\Phi/\Phi_{0})$ and the velocity amplitude $v_0$ depends on  $\delta\Phi$ as well as on the particle's first-neighbors hopping amplitude. The above expression fits accurately the velocity dependence on $\phi$, as shown by the solid line in Fig.\ref{fig5}. For $\delta\Phi=1$, maximum positive (negative) speed is reached at $\phi_1=\delta\Phi \cos{\phi_1} \approx 0.739$  ($\phi_2=\pi-\phi_1 \approx 2.402$). 

\begin{figure}
    \centering
    \includegraphics[width=\linewidth]{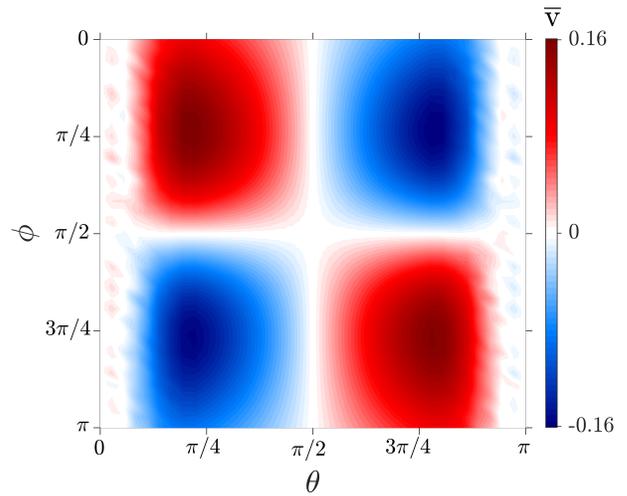}
    \caption{Density plot of the centroid's velocity in the plane $\phi$ (AC field phase) vs. $\theta$ (quantum gate). We consider  the resonance condition $\omega=\omega_{B}$, with $\omega=\Phi_{0}$, $m=100$ and $\Delta\Phi=\Phi+0=2\pi/m$. Notice that the centroid's drift velocity can be controlled by tuning either the AC field phase or the quantum gate parameter.}
    \label{fig6}
\end{figure}

The effective hopping of the quantum walker along the chain is directly influenced by the quantum gate parameter $\theta$. 
We extended our numerical experiments in order to offer a full diagram presenting the dependence of the wavepacket unidirectional centroid's velocity at resonance regime as a function of both $\theta$ vs. $\phi$. Our results are shown as a density plot in Fig. \ref{fig6}.
For $\theta = 0$ and $\pi$, the wavepacket spreads symmetrically, resulting in a stationary centroid. On the other hand, the wavepacket remains trapped around the initial position for the Pauli-X gate  $\theta =\pi/2$.
Extremal drift velocities (maximum and minimum) are achieved at  finite $\theta$ values. However, these extremal are displaced towards $\theta=0$ (or $\theta=\pi$) when smaller phase increments $\Phi_0$ are considered. Therefore,  the drift velocity of the wavepacket centroid can be controlled by tuning either the AC field phase or the gate parameter, thus opening a nice perspective for the manipulation of quantum walkers.

\section{Concluding remarks}\label{sec:conclusions}

In this work, we explored the dynamics of a quantum walker subjected to a superposition of emulated static and harmonic electric fields. The artificial constant field is related to a phase with a linear dependence on position, while the effective time-dependent field comes from a phase which exhibits a harmonic modulation.  We show different quantum walk dynamics by tuning the resulting electric field characteristics. 

Under the action solely of a weak artificial constant field, the quantum walk develops quasi-stationary Bloch-like oscillations. Super-Bloch oscillations, with large amplitude and low frequency $\omega_{SBO}$, can be achieved when the frequency of the harmonic phase field is closely tuned to the typical Bloch-oscillations frequency. We demonstrated that this low-frequency component on the quantum walk dynamics is predominant whenever the detuning frequency $\Delta\omega = \omega-\omega_B$ is below a characteristic fraction of $\omega_B$.

At exact resonance conditions, $\omega=\omega_B$, the quantum walk develops a unidirectional motion. Its average velocity and direction was shown to depend both on the field characteristics (amplitudes and phase) and the specific quantum coin operator implemented. In particular,  the dependence of the wavepacket velocity on the field characteristics was shown to be well captured by a continuous time approach. The coin operator parameter controls the range of possible values for the average velocity. Considering that discrete-time quantum walks can be implemented in several physical platforms, such as optical lattices, quantum cavities and Bose-Einstein condensates, the present results show distinct ways to manipulate and drive quantum walkers by tuning either network characteristics governing the specific coin operator or the phase field parameters.

\section{Acknowledgments}

This work was partially supported by CNPq (The Brazilian National Council for Scientific and Technological Development), CAPES (Federal Brazilian Agency) and FAPEAL (Alagoas State Agency).

\bibliographystyle{nature}
\bibliography{referencias}

\end{document}